\documentclass[11pt]{article}
\usepackage{amsmath,amssymb,epsfig,cite}
\advance\hoffset by -1.1truecm
\advance\voffset by -1.0truecm
\newcommand{\be}{\begin{equation}}
\newcommand{\ee}{\end{equation}}
\newcommand{\bea}{\begin{eqnarray}}
\newcommand{\eea}{\end{eqnarray}}

\numberwithin{equation}{section}

\newcounter{appendice}

\begin{document}

 \title{
 \begin{flushright}
 \end{flushright}
\vspace{2.2cm}\begin{flushleft} \bf{Noncommutative Vortices and Instantons from Generalized Bose Operators
\linethickness{.05cm}\line(1,0){433}
}\end{flushleft}}
\author{\bf{Nirmalendu Acharyya${^a}$\footnote{nirmalendu@cts.iisc.ernet.in},\,  Nitin Chandra$^{a}$\footnote{nitin@cts.iisc.ernet.in} \,and Sachindeo Vaidya$^{a,b}$\footnote{vaidya@cts.iisc.ernet.in}} \\ 
\begin{small}{\it $^{a}$Centre for High Energy Physics, Indian Institute of Science, Bangalore 560012, India}
\end{small} \\
\begin{small}{\it $^{b}$Department of Physics, McGill University, Montr\'eal, QC, Canada H3A 2T8}\end{small}
}
\date{\empty}

\maketitle

\begin{abstract}
Generalized Bose operators correspond to reducible representations of the harmonic oscillator algebra. 
We demonstrate their relevance in the construction of topologically non-trivial solutions in 
noncommutative gauge theories, focusing our attention to flux tubes, vortices, and instantons. Our method 
provides a simple new relation between the topological charge and the number of times the basic 
irreducible representation occurs in the reducible representation underlying the generalized Bose 
operator. When used in conjunction with the noncommutative ADHM construction, we find that these 
new instantons are in general not unitarily equivalent to the ones currently known in literature.

\end{abstract}


\tableofcontents

\section{Introduction}

Quantum field theories on noncommutative space-times are of considerable interest for a variety 
of reasons: they can provide a self-consistent deformation of ordinary quantum field theories at 
small distances, yielding non-locality \cite{Bahns:2002vm,Bahns:2003vb,Bahns:2004fc},
 or create a framework for finite truncation of quantum 
field theories while preserving symmetries \cite{Grosse:1995ar,Grosse:1995pr,Grosse:1995jt,Grosse:1996mz,Baez:1998he,Balachandran:1999hx}. Arguments combining quantum uncertainties 
with classical gravity also provide an alternative motivation to study these theories 
\cite{Doplicher:1994tu}, and the emergence of noncommutative field theories in string theory 
\cite{Dhar:1994ib,Douglas:1997fm,Seiberg:1999vs}  have provided a considerable impetus to their 
investigation. 

Detailed investigations of noncommutative gauge theories have led to the discovery of localized static 
classical solutions in noncommutative spaces \cite{Gopakumar:2000zd,Aganagic:2000mh,Gross:2000ss,Lechtenfeld:2001aw, Grisaru:2003gt, Hamanaka:2005cd,Harvey:2000jt,Harvey:2000jb,Nekrasov:2000ih}. Among the models of gauge theories in noncommutative spaces, one of the simplest is the 
Abelian-Higgs model possessing vortex--like solutions \cite{Jatkar:2000ei, Bak:2000ac, Bak:2000im, Lozano:2000qf}.
Another interesting class of solutions in noncommutative euclidean space are instantons in $U(N)$ Yang-Mills theories. 
Nekrasov and Schwarz developed a generalization of ADHM construction as given in \cite{Atiyah:1978ss} to find these noncommutative instantons \cite{Nekrasov:1998ss}. 
The $U(N)$ Yang-Mills instantons in $\mathbb{R}_{NC}^4 $ and  $\mathbb{R}_{NC}^2 \times\mathbb{R}_{C}^2 $ were further studied in 
\cite{Furuuchi:1999kv,Kim:2000msa,Chu:2001cx,Kim:2001ai, Lechtenfeld:2001ie, FrancoSollova:2002nn,Furuuchi:2000vc}. 
Pedagogical reviews can be found in \cite{Nekrasov:2000zz, Sako:2010zza}. Apart 
from these, there are multitudes of other solutions in noncommutative gauge fields like merons \cite{FrancoSollova:2002nn}, flux tubes 
\cite{Polychronakos:2000zm}, monopoles \cite{Gross:2000wc}, dyons \cite{Hashimoto:1999zw}, 
skyrmions \cite{Ezawa:2005cj}, false vacuum bubbles \cite{Bak:2000ac}, to name just a few. 

In this article we present a new construction of such topological objects, based on an analysis of the 
{\it reducible} representations of the standard harmonic oscillator algebra. Our method gives rise to 
new instanton solutions (i.e. not gauge equivalent to the known ones), and in the process provides a 
simple interpretation for the instanton number: it simply ``counts" the number of copies of the basic 
irreducible representation. 
 
Our construction relies on operators called generalized Bose operators \cite{bg,katriel} which provide
 an explicit realization of the reducible representations of the oscillator algebra, and are well-known 
in the quantum optics literature (see for examples \cite{katriel2, buzek}). As a warm-up, 
we first study the significance of generalized Bose operators in constructing fluxes and vortices with higher 
winding numbers, and then discuss instanton solutions in noncommutative YM theories.
 
The article is organized as follows. We start with a brief review of generalized 
Bose operators and its representations in section \ref{bg_review}. In section \ref{static_soln}, 
we discuss the flux tube solutions of \cite{Polychronakos:2000zm} in the language of 
generalized Bose operators and then we go on to show the relevance of these operators in 
noncommutative Nielsen-Olesen vortices. Section \ref{instantons} discusses the noncommutative 
instantons. using the generalized Bose operator in conjunction with the ADHM construction, We construct 
a class of new instantons and compute 
their topological charges. Our conclusions are presented in section \ref{conclusion}.

\section{Generalized Bose Operators -- A Brief Review}\label{bg_review}

Brandt and Greenberg \cite{bg} give a construction of generalized Bose operators that 
change the number of quanta of the standard Bose operator $a$ by 2 (or more generally 
by a positive integer $k$). We briefly recall their construction in this section.

Consider the infinite-dimensional Hilbert space $\mathcal{H}$ spanned by 
a complete orthonormal basis $\{|n \rangle, n=0,1,\cdots, \infty \}$ labeled by a 
non-negative integer $n$. Vectors in $\mathcal{H}$ are of the form $|\psi \rangle = \sum_n c_n |n \rangle, c_n \in {\mathbb C} \quad \forall n$ such that $\sum_n |c_n|^2 < \infty$.

The standard bosonic annihilation operator $a$ acts on this basis as
\begin{equation}
 a |n\rangle = n^{\frac{1}{2}} |n-1\rangle, \quad \forall n\geq 1 \quad \mathrm{and} 
 \quad a|0\rangle=0
\end{equation}
The annihilation operator is unbounded, and hence comes with a domain of definition:
\begin{equation}
\mathcal{D}_a = \{ \sum_n c_n |n \rangle \arrowvert \sum_n n |c_n|^2 < \infty \}
\end{equation}
Its adjoint $a^\dagger$ satisfies
\begin{equation}
a^\dagger |n\rangle = (n+1)^{\frac{1}{2}} |n+1\rangle, \quad \forall n \geq 0
\end{equation}
and the closure of its domain is also $\mathcal{D}_a$.

The {\it number} operator $N \equiv a^\dagger a$ has as its domain of definition $\mathcal{D}_N$, where
\begin{equation}
\mathcal{D}_N = \{ \sum_n c_n |n \rangle, \quad \arrowvert \sum_n n^2 |c_n|^2 <\infty \}.
\end{equation}
The basis vectors $\{|n\rangle\}$ are eigenstates of $N$:
\begin{equation}
N|n\rangle = n |n\rangle .
\end{equation}
On $\mathcal{D}_N$, the operators $a$ and $a^\dagger$ satisfy
\begin{equation}
[a,a^\dagger]=1.
\label{basiccomm}
\end{equation}
While $N$ counts the number of quanta in a state, the $a$ and $a^\dagger$ 
destroy and create respectively a single quantum. Thus $(a,\mathcal{H})$ is a representation of the oscillator 
algebra (\ref{basiccomm}). It is also the unique (up to unitary equivalence) irreducible representation of this 
algebra (see for example \cite{kirillov}).

The Hilbert space $\mathcal{H}$ can split into two disjoint subspaces 
$\mathcal{H}_+=\{\sum c_{2n}|2n\rangle  \in \mathcal{H}\}$ and $\mathcal{H}_- 
=\{\sum c_{2n+1}|2n+1\rangle  \in \mathcal{H}\}$ : $\mathcal{H}=\mathcal{H}_+\oplus\mathcal{H}_- $. The (projection) operators
\begin{equation}
\Lambda_+ = \sum_{n=0}^\infty |2n\rangle\langle2n|, \quad \Lambda_- = \sum_{n=0}^\infty |2n+1 \rangle \langle 2n+1|
\end{equation}
project onto the subspaces $\mathcal{H}_\pm$. On the subspaces $\mathcal{H}_{\pm}$,  the operators $b_\pm$ and its adjoint  
$b^\dagger_\pm$ can be defined as 
\begin{eqnarray}
\begin{array}{l l l l}
 b_+ |2n\rangle = n^{\frac{1}{2}} |2n-2\rangle, \quad   b^\dagger_+ |2n\rangle = (n+1)^{\frac{1}{2}} |2n+2\rangle,   \quad  b_+ |0\rangle =0,  \\
b_-|2n+1\rangle = n^{\frac{1}{2}} |2n-1\rangle,  \quad b^\dagger_-|2n+1\rangle = (n+1)^{\frac{1}{2}} |2n+3\rangle, \quad b_- |1\rangle =0
\end{array}
\end{eqnarray}
with domain of closure $\mathcal{D}_a \cap \mathcal{H}_{\pm}$. 

On the domain $\mathcal{D}_N \cap \mathcal{H}_{\pm}$ we have $[b_\pm,b_\pm^\dagger ]=1$.
Thus $(b_-,\mathcal{H}_-)$, $(b_+,\mathcal{H}_+)$ and $(a,\mathcal{H})$ are isomorphic to 
each other. In other words, there exist unitary operators $U_\pm$ such that $U_\pm b_\pm 
U_\pm^\dagger = a$.

Using the projection operators $\Lambda_\pm$, one can define an operator $b$ 
\begin{equation}
 b=b_+\Lambda_+ +b_-\Lambda_- 
\label{irreducible_decomp}
\end{equation}
on $\mathcal{H}$ whose action  on the basis vectors $|n\rangle$ is
\begin{equation}
 b|2n\rangle = n^\frac{1}{2} |2n-2\rangle, \quad b|2n+1\rangle = n^\frac{1}{2} |2n-1\rangle
\label{action_b}.
\end{equation}
Notice that both $|0\rangle$ and $|1\rangle$ are annihilated by $b$.

The operator $b$ satisfies the commutation relation $[N, b] =-2b$ and  a new number operator can be defined 
as $M=b^\dagger b= \frac{1}{2}\left(N-\Lambda_-\right)$ which has the states $|n\rangle$  as eigenstates but each eigenvalue 
is two-fold degenerate. We can denote these eigenvalues  by $m_n=\frac{1}{2} (n-\lambda_-) $ where $\lambda_- (= \langle 
n|\Lambda_-|n\rangle)$ takes values $0$ and $1$ for even and odd $n$'s respectively. Then (\ref{action_b}) can be rewritten as
\begin{equation} \label{action_b_1}
 b|n\rangle=m_n^\frac{1}{2}|n-2\rangle \quad \mathrm{and} \quad  b^\dagger|n\rangle=(m_n+1)^\frac{1}{2}|n+2\rangle.
\end{equation}
The operator $b$ has domain of closure $\mathcal{D}_a$ and satisfy $[b,b^\dagger] =1$ in the domain $\mathcal{D}_N$ and thus ($b,
\mathcal{H}$) forms a reducible representation of the algebra $[a,a^\dagger]=1$ having (\ref{irreducible_decomp}) as its irreducible 
decomposition.

The above can be generalized to construct an operator $b^{(k)}$ which lowers a state 
$|n\rangle$ by $k-$steps. We start by defining projection operators $\Lambda_i$ by
\begin{equation}
\Lambda_i = \sum_{n=0}^\infty |k n +i\rangle \langle k n + i |, \quad i = 0,1, \cdots k-1.
\end{equation} 
that project onto subspaces $\mathcal{H}_i =\{\sum_n c_{kn+i} |k n +i\rangle \}$.
In each subspace $\mathcal{H}_i$, one can define operators $b_{i}$ and their adjoints 
$b^\dagger_{i}$ that satisfy $[b_{i}, b^\dagger_{i}]=1$ and hence correspond to the UIR of the oscillator algebra. A reducible representation is given by
\begin{equation}
 b^{(k)} = \sum_{i=0}^{k-1} b_i \Lambda_i, \quad b_i |kn+i\rangle = \sqrt{n} |kn+i-k\rangle, 
 \quad \mathcal{H}=\sum_{i=0}^{k-1}\mathcal{H}_i
\label{b_red}
\end{equation}
with $[b^{(k)},b^{(k)\dagger}]=1$. Again,$(b_i,\mathcal{H}_i)$ is isomorphic to $(a,\mathcal{H})$ and $(b^{(k)},\mathcal{H})$ 
forms a reducible representation of the algebra $[a,a^\dagger]=1$.

The equations (\ref{irreducible_decomp})--(\ref{action_b_1}) represent the case $k=2$, the simplest non-trivial example of this construction.
Henceforth we will use $b$ for $b^{(2)}$. An explicit expression for $b$ is \cite{katriel} 
\begin{equation}
 b= \frac{1}{\sqrt{2}} \left( a\frac{1}{\sqrt{N}}a\Lambda_+ +a\frac{1}{\sqrt{N+1}}a\Lambda_-\right)
\label{expression_b}
\end{equation}

Before we end this section, let us point out a minor generalization of the 
Brandt-Greenberg construction. Under any unitary transformation $U$ that acts on $b_+$ as 
$b_+ \rightarrow U b_+ U^\dagger$, the fundamental relation $[b_+,b_+^\dagger]=1$ is 
unchanged. In particular, if we choose $U \equiv U_+(z_+) = e^{z_+ b_+^\dagger - \bar{z}_+ b_+}$, 
then we find that 
\begin{equation}
b_+(z_+) \equiv U_+(z_+) b_+ U_+^\dagger(z_+) = b_+ - z_+,
\end{equation}
i.e. we get the ``translated" annihilation operator. We can construct a reducible representation using 
$b_+(z_+)$ and $b_-(z_-)$ (defined similarly) as
\begin{equation}
b(z_+,z_-) = b_+(z_+) \Lambda_+ + b_-(z_-) \Lambda_- = b_+ \Lambda_+ + b_-\Lambda_- -z_+\Lambda_+ -z_-\Lambda_-.
\end{equation}
Here $b$  gets translated by different amounts in different subspaces $\mathcal{H}_{\pm}$ and the ``translated'' operator 
$b(z_+,z_-)$ is unitarily related to $b$ as
\begin{equation}
b(z_+,z_-) =U(z_+,z_-)b U^\dagger(z_+,z_-),  \quad U (z_+,z_-) = U_{+}(z_+)\Lambda_{+}+U_{-}(z_-)\Lambda_{-}.
\label{translation_operator}
\end{equation}

More generally, using (\ref{b_red}) we can write
\begin{equation}
b^{(k)}(z_0,z_1,\cdots,z_{k-1}) = b^{(k)} - \sum_{i=0}^{k-1} z_i \Lambda_i
\end{equation}
and the unitary operator is $U(z_0,z_1,\cdots,z_{k-1}) = \sum_{i=0}^{k-1} U_{i}(z_i) \Lambda_i $.
Though minor, this generalization will play a role in the construction of noncommutative 
multi-instantons.

There exist other possibilities as well. For example, choosing $U(\lambda_+) = 
e^{\frac{\lambda_+}{2}(b_+^2 -b_+^{\dagger 2})}$ gives
\begin{equation}
b_+(\lambda_+) = b_+ \cosh \lambda_+ + b_+^\dagger \sinh \lambda_+.
\end{equation}
The above is the well known squeezed annihilation operator. Using $b_+(\lambda_+)$ and $b_-(\lambda_-)$, a reducible representation may be constructed:
\begin{equation}
b(\lambda_+,\lambda_-) = b_+(\lambda_+)\Lambda_+ + b_-(\lambda_-) \Lambda_-
\end{equation}
and more generally,
\begin{equation} \label{reducible_squeeze}
b(\lambda_0,\lambda_1, \cdots,\lambda_{k-1}) = \sum_{i=0}^{k-1} b_{i}(\lambda_i)\Lambda_i
\end{equation}

\section{Static Solutions In Noncommutative Gauge Theories }\label{static_soln}

We are interested in exploring the relevance of the generalized Bose operators $b^{(k)}$ 
in noncommutative field theories. Let us start by considering two simple situations:
\begin{itemize}
\item the flux tube solution in $(3+1)-$dimensional pure gauge theory 
\item the vortex solution in $(2+1)-$dimensional abelian Higgs model.
\end{itemize}

\subsection{Flux Tube Solution In Noncommutative Gauge Theories}{\label{sec:flux_sec}}
Consider pure $U(1)$ gauge theory in $(3+1)$-dimensional spacetime with only 
spatial noncommutativity. This theory incorporates magnetic flux tube solutions which  are important 
in the context of monopoles and strings.

We are interested in the non-trivial solutions of the static equation 
of motion. These solutions do not  possess a smooth 
$\theta\rightarrow 0$ limit \cite{Polychronakos:2000zm}, implying that these configurations have 
no commutative counterpart, and the origin of this effect is entirely due  the noncommutativity of 
the underlying space.   

Since the time coordinate commutes with spatial coordinates in this model, we can, by 
an appropriate choice of axes, choose the noncommutativity to have the form
\begin{equation}
 [\hat{x}^1,\hat{x}^2]=i\theta, \quad [\hat{x}^1,\hat{x}^3]=0, \quad [\hat{x}^2,\hat{x}^3]=0, 
\end{equation}
so that only the $\hat{x}^1-\hat{x}^2$ plane is noncommutative. 

In noncommutative space, ``functions" are elements of the noncommutative algebra 
(i.e. operators) generated by the operators $\hat{x}^i$.  We will work directly with such operators.
Derivatives in the $\hat{x}^1$ and $\hat{x}^2$ 
directions are defined via the adjoint action
\begin{equation}
 \partial_{x^1}f=\frac{i}{\theta}\left[\hat{x}^2,f\right], \quad \partial_{x^2}f=-\frac{i}{\theta}\left[\hat{x}^1,f\right].
\label{derivative}
\end{equation}
while the derivatives in the $\hat{x}^3$ and $t$ directions are the same as in the commutative case.

We can  define  a set of complex (noncommuting) variables $z$ and $\bar{z}$ and a set of creation-annihilation operators as
\begin{equation}
 z= \frac{1}{\sqrt{2}}(\hat{x}^1+i\hat{x}^2),  \quad  \bar{z}= \frac{1}{\sqrt{2}}(\hat{x}^1-i\hat{x}^2), \quad a=\frac{1}{\sqrt{\theta}}z,
\quad  a^\dagger=\frac{1}{\sqrt{\theta}}\bar{z}
\end{equation}
which satisfy $[a,a^\dagger]=1$. With this convention, the derivatives with respect to the complex coordinates are 
given as
\begin{equation}
 \partial_{z}f=-\frac{1}{\sqrt{\theta}}\left[a^\dagger,f\right], \quad \partial_{\bar{z}}f=\frac{1}{\sqrt{\theta}}
\left[a,f\right].
\end{equation}

 Ordinary products become operator products and the integration on $x^1-x^2$ plane is replaced by trace over Fock space $\mathcal{H}$:
 $ \int dx^1 dx^2 f(x^1,x^2) \rightarrow 2\pi\theta \mathrm{Tr}_{\mathcal{H}}\hat{f}(\hat{x}^1,\hat{x}^2)$.

Taking $A_{\mu}$ as anti-hermitian operator valued functions, the action of $U(1)$ pure gauge theory   
in the above noncommutative space is
\begin{equation}
 S= -\frac{\pi\theta}{2g^2}\int dx^3 dt \mathrm{Tr} \lbrace F^2_{\mu\nu}\rbrace
\end{equation}
where the field strength is $ F_{\mu\nu}=\partial_{\mu} A_{\nu} - \partial_{\nu} A_{\mu}+[A_{\mu},A_{\nu}]$.

We can define noncommutative covariant derivative as 
 \begin{equation}
  \mathcal{D}_{\mu} = i \theta_{\mu\nu} x^\nu +  A_{\mu}, \quad (A_{\mu})^\dagger = -A_{\mu}.
 \end{equation}
Noncommutative gauge theory has an alternative (and possibly more natural) formulation in terms 
of $\mathcal{D}_{\mu}$ rather than $A_{\mu}$. In terms of $\mathcal{D}_{\mu}$, one works with 
the action 
\begin{equation}
 \hat{S} = -\frac{\pi\theta}{2g^2}\int dx^3 dt \mathrm{Tr} \lbrace\hat{F}^2_{\mu\nu}\rbrace, \quad\quad \mathrm{where} 
\quad \hat{F}_{\mu\nu}=[\mathcal{D}_{\mu},\mathcal{D}_{\nu}],
\end{equation}
which is (classically) equivalent to the $S$: $ \hat{S}=S +\frac{\pi}{g^2 \theta}\int dt dx^3 \left(\mathrm{Tr}I-i\theta F_{12}\right)=S+$Constant term $+$ boundary term. Thus both $\hat{S}$ 
and $S$ give the same equations of motion.

For static, magnetic configurations $(\partial_t=0 = A_{0})$ and with the choice 
$\partial_{3}A_{i}=0,\mathcal{D}_{3}=0$, we have
\begin{equation}
 [\mathcal{D},[\mathcal{\bar{D}},\mathcal{D}]]=0, \quad \quad \mathrm{with} \quad \mathcal{D}=\frac{1}{\sqrt{2}}( \mathcal{D}_{1}+i\mathcal{D}_{2}), 
\quad \mathcal{\bar{D}}=\frac{1}{\sqrt{2}}( \mathcal{D}_{1}-i\mathcal{D}_{2}).
\label{eom_flux}
\end{equation}
It is easy to check that $\mathcal{D}=\frac{a}{\sqrt{\theta}}$ is a solution of (\ref{eom_flux}), which is the vacuum configuration and has field strength 
$F_{\mu\nu} =0$. We can construct solutions about this vacuum by taking a rotationally invariant ansatz  $D=af(N)$ and it 
can be shown that there exists a  solution of (\ref{eom_flux})
\begin{equation}
 D= \frac{a}{\sqrt{\theta}}\sqrt{\frac{N-n_0}{N}}\sum_{n=n_0}^\infty |n\rangle \langle n|, \quad N=a^\dagger a, \quad n_0 =0,1,2.... 
\end{equation}
with a localized flux representing classical static circular magnetic flux tubes in $x^3$ direction centered about origin of the $(x^1,x^2)$ 
plane with $n_0$ related to its radius. The choice $n_0 =0$ corresponds to the vacuum configuration and has zero magnetic flux.

By virtue of the relation $[b,b^\dagger]=1$, $D=\frac{b}{\sqrt{\theta}}$ is also a solution of (\ref{eom_flux}).  
Again, we can start with the ansatz $D= G(N) b$ to construct the following solution of (\ref{eom_flux}) 
\begin{equation}
 D= \frac{b}{\sqrt{\theta}}\sqrt{\frac{N-n_+\Lambda_+ -n_- \Lambda_-}{2M}}
\left(\sum_{n=n_+}^\infty |n\rangle \langle n|\Lambda_+ +\sum_{n=n_-}^\infty |n\rangle \langle n|\Lambda_-\right), \quad n_\pm=0,1,2...
\end{equation}
This solution can be re-written in the form $D= \sum g(n)|n\rangle \langle n+2|$ :
\begin{eqnarray}
 D= \frac{1}{\sqrt{\theta}}\left(\sum_{n=n_+}^\infty \lambda_+\sqrt{\frac{n-n_++2}{2}} |n\rangle \langle n+2|+ 
\sum_{n=n_-}^\infty \lambda_-\sqrt{\frac{n-n_-+2}{2}} |n\rangle \langle n+2|\right)
\end{eqnarray}
which gives $g(2n)= \sqrt{n-n_+^{\prime} +1}$ for $ n\geq n_+^{\prime}$ and $g(2n+1) =\sqrt{n-n_-^{\prime} +1}$ 
for  $  n\geq n_-^{\prime}$. This solutuion is same as the higher moment solution obtained in \cite{Polychronakos:2000zm} starting with the ansatz $D=F(N)a^2$. 
Here  $\lambda_+ =\cos^2 \left(\frac{n\pi}{2}\right)$ and $\lambda_-= \sin^2 \left(\frac{n\pi}{2}\right)$ are eigen functions of $\Lambda_\pm$ and 
$n_\pm^{\prime}$ is defined as $n_+^{\prime}= \frac{n_+}{2},n_-^{\prime}= \frac{n_--1}{2}$.
Again, for the choice $n_+=0$ and $ n_-=1$, this solution reduces to $D=\frac{b}{\sqrt{\theta}}$.

Furthermore,  $D=\frac{b^{(k)}}{\sqrt{\theta}}$ for $k\geq 2$ also solves (\ref{eom_flux}) and with the ansatz $D= f(n)b^{(k)} $, the solution 
 \begin{equation}
 D= \frac{b^{(k)}}{\sqrt{\theta}}\sqrt{\frac{N-\sum_{i=1}^k n_i\Lambda_i}{2M^{(k)}}}
\sum_{i=1}^k\sum_{n=n_i}^\infty |n\rangle \langle n|\Lambda_i,\quad M^{(k)}= b^{(k)\dagger}b^{(k)},  \quad n_i=0,1,2...
\end{equation}
can be constructed, which is also same as the higher moment solution obtained from the ansatz $D=g(n)a^k$ in \cite{Polychronakos:2000zm}.
 These solutions with the generalized Bose operator represent magnetic flux tubes, nonperturbative in $\theta$, with localised flux   and varying radial profile determined by the set 
$\lbrace n_i\rbrace$. 
This shows that there already exist  certain solutions of noncommutative gauge theory which can be re-written in terms of the generalized Bose operators. 
This fact shows the importance of generalized Bose operators and motivates us to seek such solutions in other noncommutative gauge theories.

\subsection{Noncommuative Abelian Higgs Model - Nielsen-Olesen Vortex Solution \label{nov}}

The abelian Higgs model in noncommutative spaces is of some interest because of its simplicity 
as a noncommutative gauge theory and the existence of vortex solutions. 
Various topologically non-trivial vortex solutions in this context have been studied in detail. 

An interesting class of vortex solutions in this 
theory is studied in  \cite{Jatkar:2000ei}, which are analogous to the Nielsen-Olesen Vortices in the commutative space.
The model is in $(2+1)-$dimensions, and consists of a complex Higgs filed $\Phi$ which is a left gauge module (the gauge fields multiply the 
complex Higgs field $\Phi$ from left and $\bar{\Phi}$ from right). Minimizing the static noncommutative energy functional,
 Bogomolnyi equations are generalized to the noncommutative space and  $1/\theta$-expansion is done in large $\theta$ limit. 
The equations are then solved order by order and the corrections to the leading order equation converge rapidly. In the large distance limit 
(which is the commutative limit in this case), the solution reduces to the Nielsen-Olesen vortex solution in ordinary (commutative) 
abelian Higgs model.

The noncommutativity is same as that in section (\ref{sec:flux_sec}), with the only difference that now the space is 
$2-$dimensional, so the direction $x^3$ is absent. The energy functional in the  static 
configuration is given by \cite{Jatkar:2000ei}
\begin{equation}
 \mathcal{E} = \mathrm{Tr} \left[\frac{1}{2} (B+ \Phi\bar{\Phi}-1)^2+\mathcal{D}_{\bar{z}}\Phi\mathcal{D}_z\bar{\Phi} +
\mathcal{T}\right]
\label{energy_func}
\end{equation}
where $\mathcal{D}$ is a covariant derivative with the gauge field $A$. The magnetic field $B$ is defined as $B=-i(\partial_z 
A_{\bar{z}}-\partial_{\bar{z}}A_z)-[A_z,A_{\bar{z}}]$ and $\mathcal{T}$ is the topological term defined as
\begin{equation}
 \mathcal{T}= \mathcal{D}_m S^m+B \quad \mathrm{where} \quad \mathcal{D}_m S^m=\partial_m S^m-i[A_m,S^m], \quad m= z \,\,{\rm or} \,\,\bar{z}.
\end{equation}
with $S^m= \frac{i}{2} \epsilon^{mn} (\Phi \mathcal{D}_n\bar{\Phi}- \mathcal{D}_n \Phi\bar{\Phi}).$ It can  be 
shown that $\mathrm{Tr}\mathcal{T}$ corresponds to the topological charge.

Our prime interest is to study the Bogomolnyi equations. Minimizing (\ref{energy_func}) we get the following operator 
equations:
 \begin{equation}
 \mathcal{D}_{\bar{z}}\Phi=0, \quad \mathcal{D}_z\bar{\Phi}=0, \quad B=1-\Phi\bar{\Phi}.
 \end{equation}
which are the noncommutative Bogomolnyi equations.
 Now we can do a $1/\theta$ expansion of the Higgs field, gauge field and the magnetic field in the large $\theta$ limit 
as 
\begin{eqnarray}
 \begin{array}{lll}
  \Phi &=& \Phi_\infty +\frac{1}{\theta}\Phi_{-1}+..... \\
     A &=& \frac{1}{\sqrt{\theta}} (A_\infty+\frac{1}{\theta}A_{-1}+.....)\\
B &=&\frac{1}{\theta} (B_\infty+\frac{1}{\theta}B_{-1}+.....).\\
 \end{array}
\end{eqnarray}
The factor of $\frac{1}{\sqrt{\theta}}$ and $\frac{1}{\theta}$ is used for scaling the variables $A$ and $B$ as they are 
are 1-form and 2-form respectively. With this expansion, we can get the leading order $O(\theta)$ Bogomolnyi equation as
\begin{equation}
 \Phi_\infty\bar{\Phi}_\infty=1.
\label{eom_lead}
\end{equation}
This equation admits a solution \cite{Witten:2000nz} 
\begin{equation}
 \Phi_\infty= \frac{1}{\sqrt{a^n a^{\dagger n}}}a^n, \quad \bar{\Phi}_\infty= (\Phi_\infty)^\dagger
\label{witten_vortex}
\end{equation}
which represents a $n$-vortex at origin. A more general solution is discussed in \cite{Jatkar:2000ei} which represents $n$ single 
vortices at $n$ different points in the noncommutative plane and (\ref{witten_vortex}) is a special case of that general solution. But for our discussion, (\ref{witten_vortex}) 
is sufficient and due to its simple form, computation and understanding becomes easier.

Next we take the $O(1)$ Bogomolnyi equation  
\begin{equation}
 [a,\Phi_\infty]=i\bar{A}_\infty \Phi_\infty
\label{eom_sublead}
\end{equation}
which can be solved to get (for details see  appendix \ref{det_gauge_field})
\begin{equation}
 \bar{A}_\infty = -i \frac{1}{\sqrt{N+1}} a\left(\sqrt{N}-\sqrt{N+n}\right), \quad {A}_\infty= (\bar{A}_\infty)^\dagger
\end{equation}
where $N$ is the number operator.

In the coherent state $|\omega\rangle $ ($a|\omega\rangle=\omega|\omega\rangle$), the expectation of the field $\Phi_\infty$ is 
\begin{equation}
 \langle\omega|\Phi_{\infty}| \omega \rangle= \omega^n \langle\omega|\frac{1}{\sqrt{a^n a^{\dagger n}}}| \omega \rangle
\quad \mathrm{with} \quad \omega=|\omega| e^{i n\varphi}.
\end{equation}
Here the phase dependence is $ e^{i n\varphi}$ which comes solely from  $\omega^n$ as the other factor
$\langle\omega|\frac{1}{\sqrt{a^n a^{\dagger n}}}$ $| \omega \rangle$ is purely real, signifying a vortex in the noncommutative plane.
The large distance behavior is given by the  large 
$\omega $ limit or equivalently  large $\langle N\rangle $ limit. The coherent state expectations in this limit becomes (for details see appendix \ref{l_d_b})
\begin{eqnarray}
 \langle\omega|\Phi_{\infty}| \omega \rangle \approx e^{i n\varphi},\quad \langle\omega|\bar{A}_\infty| \omega \rangle
\approx  i\frac{n}{2\bar{\omega}},\quad \langle\omega|A_\infty| \omega \rangle\approx - i\frac{n}{2\omega},
\label{large_distance1}
\end{eqnarray}
which  is exactly like the commutative $n$-Nielsen-Olesen Vortex.

It is interesting to note that the leading order magnetic field is $B_\infty= n|0\rangle\langle 0|$, which means the magnetic 
field of the solution is localized and the magnetic fluxes are confined. This problem is qualitatively similar to flux 
tube problem in Section \ref{sec:flux_sec}, where we saw that the theory has solutions 
involving the generalized Bose operator $b$. This fact stimulates us to seek vortex solutions 
with operator $b$ in the Higgs model, i.e., a solution to (\ref{eom_lead}) involving the $b$'s
 and a corresponding new gauge field  ${A}_\infty$ .  With this motivation in mind one can  easily check 
\begin{equation}
 \Phi^{new}_\infty= \frac{1}{\sqrt{b^n b^{\dagger n}}}b^n, \quad \bar{\Phi}_\infty= (\Phi_\infty)^\dagger
\label{new_vortex}
\end{equation}
satisfies (\ref{eom_lead}) and using this, as before, we can find (details in appendix \ref{det_gauge_field})
 \begin{equation}
 \bar{A}^{new}_\infty=-i\left(a- \frac{1}{\sqrt{b^n b^{\dagger n}}}b^nab^{\dagger n}\frac{1}{\sqrt{b^n b^{\dagger n}}}\right), 
\quad {A}^{new}_\infty= (\bar{A}_\infty)^\dagger.
\label{new_gauge_field}
\end{equation}
The expectation value in the coherent state $|\omega\rangle $ (eigenstate of $a$) gives a phase dependence of $e^{i 2n\varphi}$ 
(as $\Phi^{new}_\infty$ can always be reduced to the form $F(N)a^{2n}$ ), a characteristic feature of $2n$ vortex in noncommutative
plane. In the  large $\omega $ limit it gives the large distance behavior:
\begin{eqnarray}
 \langle\omega|\Phi^{new}_{\infty}| \omega \rangle \approx e^{i 2n\varphi},\quad \langle\omega|\bar{A}^{new}_\infty| \omega \rangle
\approx  i\frac{n}{\bar{\omega}},\quad \langle\omega|A^{new}_\infty| \omega \rangle\approx - i\frac{n}{\omega}
\label{large_distance2}
\end{eqnarray}
which is exactly the commutative $2n$ Nielsen-Olesen vortex.

Now we need to compare the new solution (\ref{new_vortex}) with the Witten's solution (\ref{witten_vortex}). For the simplicity of 
expression and better understanding of the underlying algebra, we take $n=1$ in (\ref{new_vortex}). Using the explicit expression of 
the generalized Bose operator $b$, the new vortex solution can be
written as  
\begin{eqnarray}
  \Phi^{new}_\infty =& \frac{1}{\sqrt{bb^\dagger}} b 
&= \frac{1}{\sqrt{M+1}} \frac{1}{\sqrt{2}} (a\frac{1}{\sqrt{N}}a \Lambda_+ + a\frac{1}{\sqrt{N+1}}a\Lambda_-)
\label{reduction1} 
\end{eqnarray}
where we know, $M=b^\dagger b= \frac{N-\Lambda_-}{2}$.
Further simplification can be done and the expression (\ref{reduction1}) for the new vortex reduces to 
\begin{equation}
 \Phi^{new}_\infty=(\frac{1}{\sqrt{N-\Lambda_-+2}}\frac{1}{\sqrt{N+1}}\Lambda_++\frac{1}{\sqrt{N-\Lambda_-+2}}\frac{1}{\sqrt{N+2}}
\Lambda_-) a^2.
\label{reduction2} 
\end{equation}
The eigenvalues of the projection operators $\Lambda_{\pm}$ are $0$ and $1$ and they never contribute simultaneously.
Owing to this fact the expression 
(\ref{reduction2}) simplifies to 
\begin{equation}
 \Phi^{new}_\infty=\frac{1}{\sqrt{N+2}}\frac{1}{\sqrt{N+1}}(\Lambda_++\Lambda_-) a^2
=\frac{1}{\sqrt{(N+1)(N+2)}} a^2
=\frac{1}{\sqrt{a^2 a^{\dagger 2}}}a^2
\end{equation}
which is same as the $n=2$  Witten's vortex. 
 This calculation can be generalized for any $n$ and it can be always shown that 
$n$-new vortex solution is same as the $2n$-Witten vortex for all $n$.
It is also easy to show that $\Phi_\infty=\frac{1}{\sqrt{(b^{(k)})^n(b^{(k)\dagger})^n}} (b^{(k)})^n $ is also a solution of (\ref{eom_lead}) 
and this solution is same as the $kn$-Witten vortex.

This shows that the vortex solutions with the generalized Bose operators are already present in the old 
solutions but are subtly hidden. It is also very interesting to see how the projectors intelligently conspire 
to make it happen.  

\section{Instantons}\label{instantons}

\subsection{Instantons In Commutative Gauge Theories}
Instantons are localized finite action solutions of the classical Euclidean field equations of a theory (for a review see \cite{Rajaraman:1982is}).
The finite action condition is satisfied only if the Lagrangian density of the theory vanishes at boundary.
This in turn can lead to different topological configurations of the field characterized by its 
``topological charge''.
For  Yang-Mills theories,  the instantons  are further classified as 
Self-Dual (SD) or Anti-Self-Dual (ASD) with their topological charges having opposite signs.
A  simple prescription to construct (anti-) self-dual instantons in the Yang-Mills theory is given in \cite{Atiyah:1978ss}.
 Let us first review this construction.

In order to describe charge $k$ instantons with gauge group $U(N)$ on $\mathbb{R}^4$ one starts with the following data:
\begin{enumerate}
\item A pair of complex hermitian vector spaces $V=\mathbb{C}^k$ and $W=\mathbb{C}^N$.
\item The operators $B_1,B_2 \in \rm{Hom}(V,V)$, $I \in \rm{Hom}(W,V)$, $J \in \rm{Hom}(V,W)$, which must obey the equations 
\begin{equation} \label{adhm_condition}
[B_1,B_1^\dagger]+[B_2,B_2^\dagger]+II^\dagger - J^\dagger J = 0, \quad
[B_1,B_2]+IJ = 0
\end{equation}
\end{enumerate}
For $z=(z_1,z_2) \in \mathbb{C}^2 \approx \mathbb{R}^4$, define an operator 
$\mathcal{D}:V\oplus V \oplus W \rightarrow V \oplus V$ as
\begin{equation} \label{d_asd}
\mathcal{D}^\dagger = \left(
\begin{array}{c}
 \tau \\
\sigma^\dagger
\end{array}
\right), \quad
\tau = \left(
\begin{array}{ccc}
 B_2 - z_2 & B_1-z_1 & I
\end{array}
\right), \quad
\sigma = \left(
\begin{array}{c}
 -B_1+z_1 \\
B_2-z_2 \\
J
\end{array}
\right)
\end{equation}
for anti-self-dual instantons and by
\begin{equation} \label{d_sd}
\mathcal{D}^\dagger = \left(
\begin{array}{c}
 \tau \\
\sigma^\dagger
\end{array}
\right), \quad
\tau = \left(
\begin{array}{ccc}
 B_2 - \bar{z}_2 & B_1+z_1 & I
\end{array}
\right), \quad
\sigma = \left(
\begin{array}{c}
 -B_1-z_1 \\
B_2-\bar{z}_2 \\
J
\end{array}
\right)
\end{equation}
for self-dual instantons. 
Given the matrices $B_1, B_2,I$ and $J$ obeying all the conditions above, 
the actual instanton solution is determined by the following rather explicit formulae:
\begin{equation} \label{gauge_field}
A_\alpha = \Psi^\dagger\partial_\alpha\Psi,
\end{equation}
($\alpha=1,2,3,4$) where $\Psi: W \rightarrow V \oplus V \oplus W$ is the normalized solution of
\begin{equation} \label{zero_mode}
 \mathcal{D}^\dagger\Psi = 0, \quad  \Psi^\dagger\Psi = 1.
\end{equation}
 Here $\partial_\alpha$ is derivative with respect to the spacetime coordinates $x_\alpha$
which are related to the $z$-coordinates as
\begin{equation}
z_1 = \frac{x_1 + ix_2}{\sqrt{2}}, \quad  \bar{z}_1 = \frac{x_1 - ix_2}{\sqrt{2}}, \quad 
z_2 = \frac{x_3 + ix_4}{\sqrt{2}}, \quad \bar{z}_2 = \frac{x_3 - ix_4}{\sqrt{2}}.
\end{equation}
For given ADHM data and the zero mode condition (\ref{zero_mode}), the following completeness
relation has to be satisfied 
\begin{equation}
\label{completeness}
\mathcal{D}\frac{1}{\mathcal{D}^\dagger \mathcal{D}}\mathcal{D}^\dagger + \Psi\Psi^\dagger = 1.
\end{equation}
 It has been shown in \cite{Kim:2000msa} that (\ref{completeness}) can be satisfied even for noncommutative spaces. 
Note that the fields $A_\alpha$ are anti-hermitian, consistent with $\Psi^\dagger\Psi=1$.
The field strength $F_{\alpha\beta}$ and its dual $\tilde{F}_{\alpha\beta}$ are given as
\begin{equation}
F_{\alpha\beta} = \partial_\alpha A_\beta - \partial_\beta A_\alpha + [A_\alpha,A_\beta],
\quad \tilde{F}_{\alpha\beta} = \frac{1}{2}\epsilon_{\alpha\beta\gamma\delta}F_{\gamma\delta}.
\end{equation}
The instantons found by the ADHM construction satisfy both the Yang-Mills equation of motion and the (anti-) self-duality condition
\begin{equation}
  D_\alpha F_{\alpha\beta} = 0, \quad \tilde{F}_{\alpha\beta} = \pm F_{\alpha\beta} 
\end{equation}
 and has topological charge  given by
\begin{equation}
 Q=-\frac{1}{16\pi^2}\int d^4x Tr(\tilde{F}_{\alpha\beta}F^{\alpha\beta}).
\end{equation}

\subsection{Noncommutative Euclidean Space}
To study instantons on a noncommutative $\mathbb{R}^4$, we will use the notation outlined below.
The 4-dimensional noncommutative euclidean space is defined by the following noncommutative coordinates:
\begin{equation}\label{commutation_x}
 [\hat{x}^\alpha , \hat{x}^\beta ] = i\theta^{\alpha\beta} ,\quad \alpha,\beta=1,2,3,4,
\end{equation}
and $\theta^{\alpha\beta}$ is a constant anti-symmetric $4\times4$ matrix.
We denote the algebra generated by these $x^\alpha$'s  by $\mathcal{A}_\theta$.
There are three distinct cases one may consider:
\begin{enumerate}
\item $\theta$ has rank 0 ($\theta^{\alpha\beta}=0 $ $\forall$ $ \alpha, \beta$).
In this case $\mathcal{A}_\theta$ is isomorphic to the algebra of functions on the ordinary $\mathbb{R}^4$.
This space may be denoted by $\mathbb{R}_C^4$.
\item $\theta$ has rank 2.
In this case $\mathcal{A}_\theta$ is the algebra of functions on the ordinary $\mathbb{R}^2$ 
times the noncommutative $\mathbb{R}^2$, which may be denoted by $\mathbb{R}_{NC}^2 \times \mathbb{R}_{C}^2$. Without 
 loss of generality, we can choose
$$\theta^{\alpha\beta} =
\left[ \begin{array}{cccc}
0 & \theta & 0 & 0 \\
-\theta &0 & 0 &0 \\
0& 0 & 0& 0\\
0&0 & 0 & 0\end{array} \right].$$ Let us define a system of complex coordinates and a set of operators as
\begin{eqnarray}\left.
\begin{array}{cccc}
 z_1=\frac{1}{\sqrt{2}}(x^2+i x^1) & \bar{z}_1=\frac{1}{\sqrt{2}}(x^2-i x^1) & z_2=\frac{1}{\sqrt{2}}(x^4+i x^3) &\bar{z}_2=\frac{1}{\sqrt{2}}(x^4-i x^3)\\
a_1=\frac{\bar{z}_{1}}{\sqrt{\theta}} 
& a_1^{\dagger}=\frac{z_{1}}{\sqrt{\theta}} 
& a_2=\frac{\bar{z}_{2}}{\sqrt{\theta}}
& a_2^{\dagger}=\frac{z_{2}}{\sqrt{\theta}}
\end{array}
\right.
\label{complex_coordinates}
\end{eqnarray}
which reduces the algebra (\ref{commutation_x}) to
\begin{equation}
[\bar{z}_{1},z_{1}]=\theta, \quad [\bar{z}_{2},z_{2}]=0, \quad [a_1,a_1^{\dagger}]=1 , \quad  [a_2,a_2^{\dagger}]=0. 
\label{commutation_z1}
\end{equation}
Here $a_1$ and $a_1^\dagger$ are like annihilation and creation operators respectively while $a_2^\dagger,a_2$ are ordinary complex numbers.
We can define a number operator by $N_1 = a_1^\dagger a_1$. 
The Fock space on which the elements of $\mathcal{A}_\theta$ act, consists of states denoted by $|n_1,z_2\rangle$.
Here $n_1$ denotes the eigenvalues of the number operator $N_1$ and can take only non-negative integral values, 
while $z_2$ can be any complex number and denotes the eigenvalues of $z_2$.

\item $\theta$ has rank 4.
In this case $\mathcal{A}_\theta$ is the noncommutative $\mathbb{R}^4$.
We choose $\theta$ to be of the form given by 
$$\theta^{\alpha\beta} =
\left[ \begin{array}{cccc}
0 & \theta^{12} & 0 & 0 \\
-\theta^{12} &0 & 0 &0 \\
0& 0 & 0 & \theta^{34}\\
0&0 & -\theta^{34} & 0\end{array} \right]=
\left[ \begin{array}{cccc}
0 & \theta & 0 & 0 \\
-\theta &0 & 0 &0 \\
0& 0 & 0& \theta\\
0&0 & -\theta & 0\end{array} \right],$$ where we have assumed $\theta^{12}=\theta^{34}=\theta$. Again, we can define a system 
of complex coordinates and a set of operators as in (\ref{complex_coordinates}) but now the algebra (\ref{commutation_x}) becomes 
\begin{equation}
[\bar{z}_{1},z_{1}]=[\bar{z}_{2},z_{2}]=\theta, \quad [a_1,a_1^{\dagger}]=[a_2,a_2^{\dagger}]=1.
\label{commutation_z2}
\end{equation}
The Fock space on which the elements of $\mathcal{A}_\theta$ act consists of states denoted by $|n_1,n_2\rangle$.
Here $n_1$ and $n_2$ denote the eigenvalues of the number operators $N_1 = a_1^\dagger a_1$ and $N_2 = a_2^\dagger a_2$ respectively 
which can take only non-negative integral values.

\end{enumerate}
As already mentioned in Section \ref{sec:flux_sec}, differentiation in the noncommutative space  is implimented as 
an adjoint as in (\ref{derivative}), while  the integration is implimented by a suitable trace.

\subsection{Noncommutative ADHM Construction}
ADHM construction for instantons has been generalized to a noncommutative space in \cite{Nekrasov:1998ss}.
The construction effectively remains same as in the commutative case, only change being the replacement of 0 in the right hand side
of the first equation of (\ref{adhm_condition}) by the noncommutative parameter $\theta$ 
for the case of $\mathbb{R}^2_{NC}\times \mathbb{R}^2_C$ and by $2\theta$ for the case of 
$\mathbb{R}^2_{NC}\times \mathbb{R}^2_{NC}$ respectively:
\begin{equation} \label{adhm_condition_r2}
[B_1,B_1^\dagger]+[B_2,B_2^\dagger]+II^\dagger-J^\dagger J = \theta, \quad
[B_1,B_2]+IJ = 0
\end{equation}
for $\mathbb{R}^2_{NC}\times \mathbb{R}^2_C$  as in \cite{Kim:2001ai} and
\begin{equation} \label{adhm_condition_r4}
[B_1,B_1^\dagger]+[B_2,B_2^\dagger]+II^\dagger-J^\dagger J = 2\theta, \quad
[B_1,B_2]+IJ = 0
\end{equation}
for $\mathbb{R}^2_{NC}\times \mathbb{R}^2_{NC}$. 
In noncommutative space, the gauge field is defined as $\hat{D}_{x^\alpha} = -i \theta_{\alpha\beta} x^\beta +  A_{x^\alpha}$ where 
$A_{x^\alpha}$ is the Yang-Mills gauge connection. Then the ansatz for gauge field is  
\begin{equation} \label{nc_gauge_field}
\hat{D}_{x^\alpha}=-i \theta_{\alpha\beta}\Psi^{\dagger}x^\beta\Psi
\end{equation}
where $\theta_{\alpha\beta}$ satisfies $\theta^{\alpha\beta}\theta_{\beta\gamma}={\delta^{\alpha}}_\gamma$.
The fields $A_{x^\alpha}$ are again anti-hermitian.
In $\mathbb{R}^2_{NC}\times \mathbb{R}^2_C$, the components of the gauge field along the commutative directions will be given by (\ref{gauge_field}), while
those along the noncommutative axes by $\hat{D}_{x^\alpha}$.

Let us first discuss the usual single anti-self-dual $U(1)$ instanton solutions ($k=1, N=1$) 
in  $\mathbb{R}^2_{NC}\times \mathbb{R}^2_{C}$ \cite{Kim:2001ai,Chu:2001cx} and in $\mathbb{R}^2_{NC}\times \mathbb{R}^2_{NC}$  
\cite{Kim:2000msa,Chu:2001cx,Furuuchi:2000vc,Sako:2010zza}.
For $k=N=1$, $B_1, B_2, I$ and $J$ are all complex numbers. 
As the noncommutative space (described by the coordinates $z$) has translational invariance, 
we can always choose the origin in such a way that $B_1$ and $B_2$  in  $\tau$ can be taken to be zero.
Thus (\ref{adhm_condition_r2}) or (\ref{adhm_condition_r4}) ensures that either  $I$ or $J$ is zero.
Without the loss of any generality we can choose $J=0$.
Now let us discuss the two cases seperately.

\subsubsection*{{\underline{$\mathbb{R}^2_{NC}\times\mathbb{R}^2_{C}$:}}}
Here we get $I= \sqrt{\theta}$ from the equation (\ref{adhm_condition_r2}).
The phase in $I$ does not effect the solution for the gauge field and hence has been taken to be zero.
The operator (\ref{d_asd}) for anti-self-dual instantons becomes 
\begin{equation}
\mathcal{D}^\dagger=
\left( \begin{array}{ccc}
-z_2 & -z_1 & \sqrt{\theta} \\
\bar{z}_1 & -\bar{z}_2 & 0
\end{array} \right)= \sqrt{\theta}\left( \begin{array}{ccc}
-a_2^\dagger & -a_1^\dagger & 1 \\
a_1 & -a_2 & 0
\end{array} \right)
\end{equation}
and its normalized zero mode solution is given by 
\begin{equation} \label{psi_form}
\Psi=
\left( \begin{array}{c}
\psi_1\\
\psi_2\\
\xi 
\end{array} \right); \quad \psi_1= a_2 \frac{1}{\sqrt{\delta \Delta}}, \quad \psi_2= a_1 \frac{1}{\sqrt{\delta \Delta}},  
\quad \xi=\sqrt{\frac{\delta}{ \Delta}}
\end{equation}
with $ \delta= a_1^\dagger a_1+a_2^\dagger a_2$ and $ \Delta=\delta+1.$
Note that the inverse of the operator $\Delta$ is well--defined, but that of $\delta$ is not since 
 $|n_1=0$, $z_2=0\rangle$ is a zero--mode of $\delta$.

\subsubsection*{{\underline{$\mathbb{R}^2_{NC}\times\mathbb{R}^2_{NC}$:}}}
In this case, (\ref{adhm_condition_r4}) gives $I=\sqrt{2\theta}$ and the operator in (\ref{d_asd}) becomes
\begin{equation}
\mathcal{D}^\dagger=
\left( \begin{array}{ccc}
-z_2 & -z_1 & \sqrt{2\theta} \\
\bar{z}_1 & -\bar{z}_2 & 0
\end{array} \right)= \sqrt{\theta}\left( \begin{array}{ccc}
-a_2^\dagger & -a_1^\dagger & \sqrt{2} \\
a_1 & -a_2 & 0
\end{array} \right).
\end{equation}
The zero mode solution is again a 3-element column matrix $\Psi$  
where we write $\psi_1 = \sqrt{\theta}a_2v$ and $\psi_2 = \sqrt{\theta}a_1v$ . Then (\ref{zero_mode}) becomes 
\begin{equation} \label{eqn_v_xi_1}
\hat{\Delta}v = \sqrt{2\theta}\xi , \quad 
 v^\dagger\hat{\Delta}v + \xi^\dagger\xi = 1
\end{equation}
where $\hat{\Delta}=\theta \hat{N} = \theta (a_1^\dagger a_1 + a_2^\dagger a_2)$.
But this operator does not have an inverse since it has a zero--mode
\begin{equation}
\hat{\Delta}|0,0\rangle = 0 
\end{equation}
and hence finding $v$ and $\xi$ is a bit tricky.
We define a shift operator $S$ such that
 \begin{eqnarray}
 SS^\dagger = 1, \quad S^\dagger S = 1- P, \quad P = |0,0 \rangle \langle 0,0 |.
\end{eqnarray}
Note that although the inverse of $\hat{\Delta}$ is not defined otherwise, 
it is well--defined when sandwiched between $S$ and $S^\dagger$.
Now we can solve for $\xi$ and $v$:
\begin{equation}
 \xi = \hat{\Phi}^{-\frac{1}{2}} S^\dagger, \quad v = \sqrt{2\theta}\frac{1}{\hat{\Delta}}\hat{\Phi}^{-\frac{1}{2}} S^\dagger, \quad \quad \quad  
\mathrm{where} \quad \hat{\Phi}= 1+ \frac{2\theta}{\hat{\Delta}}=1+\frac{2}{\hat{N}}
\end{equation}
which satisfy (\ref{eqn_v_xi_1}).

Thus we get
\begin{equation} 
 \psi_1=\sqrt{\theta}a_2 \sqrt{2\theta} \frac{1}{\hat{\Delta}} \hat{\Phi}^{-\frac{1}{2}} S^\dagger, \quad 
\psi_2=\sqrt{\theta}a_1 \sqrt{2\theta} \frac{1}{\hat{\Delta}} \hat{\Phi}^{-\frac{1}{2}} S^\dagger, \quad 
\xi = \hat{\Phi}^{-\frac{1}{2}} S^\dagger.
\end{equation}
We can define  the components of the gauge field in terms of the complex coordinates as
\begin{equation}
\hat{D}_1=\hat{D}_{x_2}+i\hat{D}_{x_1}, \quad \hat{D}_2= \hat{D}_{x_4}+i\hat{D}_{x_3}, \quad 
\hat{D}_{\bar{1}}=\hat{D}_{x_2}-i\hat{D}_{x_1}, \quad \hat{D}_{\bar{2}}=\hat{D}_{x_4}-i\hat{D}_{x_3}.
\end{equation}
Then the ansatz (\ref{nc_gauge_field}) translates into 
\begin{equation} \label{A_a}
\hat{D}_a = \frac{1}{\theta}\Psi^\dagger \bar{z}_a\Psi, \quad  \hat{D}_{\bar{a}} = -\frac{1}{\theta}\Psi^\dagger z_a\Psi= -\hat{D}_a^\dagger .
\end{equation}
The solution for  $k=1$ $U(1)$ ASD instanton becomes
\begin{equation}\label{usual_instanton}
\hat{D}_{a} = \frac{1}{\sqrt{\theta}} S\hat{\Phi}^{-\frac{1}{2}} a_a\hat{\Phi}^{\frac{1}{2}} S^\dagger, \quad
\hat{D}_{\bar{a}} = -\frac{1}{\sqrt{\theta}} S\hat{\Phi}^{\frac{1}{2}} a^\dagger_a \hat{\Phi}^{-\frac{1}{2}} S^\dagger
\end{equation}
where the shift operator $S$,  written explicitly, is
\begin{equation} \left.
\begin{array}{c}
 S=  1- \displaystyle{\sum_{n'_1=0}^\infty} |n'_1 , 0\rangle \langle n'_1 , 0| \left(1- \frac{1}{\sqrt{N_1+1}}a_1\right) \\
 S^\dagger=  1-  \displaystyle{\sum_{n'_1=0}^\infty}\left(1- a_1^\dagger\frac{1}{\sqrt{N_1+1}}\right) |n'_1 , 0\rangle \langle n'_1 , 0|
\end{array} \right\rbrace.
\end{equation}
In our notation of complex coordinates, the field strength is given by 
\begin{equation}
F_{a\bar{a}} = \frac{1}{\theta} + \left[ \hat{D}_a, \hat{D}_{\bar{a}} \right], \quad
F_{a\bar{b}} = \left[ \hat{D}_a, \hat{D}_{\bar{b}} \right];\quad   a,b = 1,2 \hspace*{.2cm}{\rm and}\hspace*{.2cm} a\neq b 
\end{equation}
and the ASD condition translates to
\begin{equation} \label{asd_r4}
F_{1\bar{1}} = - F_{2\bar{2}}, \quad F_{12} = F_{\bar{1}\bar{2}} = 0.
\end{equation}
The equations of motion in the $\mathbb{R}^2_{NC}\times\mathbb{R}^2_{NC}$ in terms of the fields $\hat{D}_{x^\alpha}$ are
\begin{equation} \label{eom_r4}
[\hat{D}_{x^\alpha},[\hat{D}_{x^\alpha},\hat{D}_{x^\beta}]]=0.
\end{equation}
The solution (\ref{usual_instanton}) satisfies both the ASD condition  and the equations of motion. 

\subsection{New Anti-Self-Dual $U(1)$ Instanton}
We can try to get new solutions for noncommutative instantons by using the  the generalized Bose operators.

\subsubsection*{{\underline{$\mathbb{R}^2_{NC}\times\mathbb{R}^2_C$:}}}
Let us first define two operators $b$ and $c$
\begin{eqnarray}\left.
\begin{array}{ccc}
&&b=\frac{1}{\sqrt{2}} a_1\frac{1}{\sqrt{N_1}}a_1 \Lambda_+ +\frac{1}{\sqrt{2}}a_1\frac{1}{\sqrt{N_1+1}}a_1 \Lambda_-, \\
&&c=\frac{1}{\sqrt{2}} a_2\frac{1}{\sqrt{N_1}}a_1 \Lambda_+ +\frac{1}{\sqrt{2}}a_2\frac{1}{\sqrt{N_1+1}}a_1 \Lambda_-
\end{array}\right.
\label{b_g_operator}
\end{eqnarray}
where $N_1=a_1^\dagger a_1, \quad \Lambda_+ =\cos ^2 (\frac{\pi N_1}{2}), \quad \Lambda_- =\sin ^2 (\frac{\pi N_1}{2}).$
Here $b$ is the  generalized operator defined in (\ref{expression_b}).
We can easily show that 
\begin{equation}
 b^\dagger b= \frac{1}{2}(N_1-\Lambda_-), \quad c^\dagger c= \frac{1}{2}N_2\frac{N_1-\Lambda_-}{N_1-1  }, \quad {\rm where}\quad N_2=a_2^\dagger a_2.
\end{equation}
The zero-mode solution  $\Psi_0$ is given by
\begin{equation}
 \psi_1= \sqrt{2} c , \quad \psi_2=\sqrt{2} b ,  \quad \xi=\delta \left(\frac{1}{\sqrt{N_1}}a_1\Lambda_+ +\frac{1}{\sqrt{N_1+1}}a_1\Lambda_-\right).
\end{equation}
But this solution is not normalized i.e. $\Psi_0^{\dagger}\Psi_0\neq1$.
Usually  the single instanton solution in $\mathbb{R}_{NC}^2 \times \mathbb{R}_{C}^2$ 
with  $U(1)$ gauge group is normalized as \cite{Chu:2001cx}
 \begin{equation}
 \Psi=\Psi_0 \frac{1}{\sqrt{\Psi_0^\dagger \Psi_0 }}.
 \end{equation}
But in our solution this technique cannot be used because in this case $\Psi_0^\dagger \Psi_0=\frac{N_1-\Lambda_-}{N_1-1}\delta (\delta-1)$ 
vanishes  when it operates on the state $|0\rangle$  and the inverse of $\Psi_0^\dagger \Psi_0$ 
does not exist.  
We fix this problem by defining 
\begin{equation}
  \Psi_{new}=\Psi_0 (1-p) \frac{1}{\sqrt{\Psi_0^\dagger \Psi_0 }}(1-p) u^\dagger, \quad \quad uu^\dagger = 1,\quad u^\dagger u =1-p
\end{equation}
where $p=|0\rangle\langle 0|$ is  a projection operator, and 
$u^\dagger$ is a shift operator projecting out the vacuum. The operator $u$ can be written as
\begin{equation}
u=\displaystyle{\sum_{n_1=0}^{\infty}}|n_1,z_2\rangle\langle n_1+1,z_2|, \quad 
u^\dagger=\displaystyle{\sum_{n_1=0}^{\infty}}|n_1+1,z_2\rangle\langle n_1,z_2|.
\end{equation}

It should be noted that the new solution in $\mathbb{R}_{NC}^2 \times \mathbb{R}_{C}^2$  is completely non-singular.

\subsubsection*{{\underline{$\mathbb{R}^2_{NC}\times\mathbb{R}^2_{NC}$:}}}
We claim the new solution to be
\begin{equation} \label{gauge_field_r4_new}
\hat{D}_{a}=\frac{1}{\sqrt{\theta}} S_{new}\hat{\Phi}_{new}^{-\frac{1}{2}} b_a\hat{\Phi}_{new}^{\frac{1}{2}} S_{new}^\dagger, 
\quad  
\hat{D}_{\bar{a}}= -\frac{1}{\sqrt{\theta}} S_{new}\hat{\Phi}_{new}^{\frac{1}{2}}b^\dagger_a \hat{\Phi}_{new}^{-\frac{1}{2}} S_{new}^\dagger
\end{equation}
where $ \hat{\Phi}_{new}=1+\frac{2}{M}$ and $M=M_1+M_2=\sum_{a=1}^2b_a^\dagger b_a$.
Here again the operator $\frac{1}{M}$ is not well-defined otherwise 
(as $M|0,0\rangle =M|0,1\rangle=M|1,0\rangle=M|1,1\rangle =0$), but is well-defined when sandwiched between $S_{new}$ and 
$S_{new}^\dagger$ (defined below) which projects out the states $|0,0\rangle$, $|1,0\rangle$, $|0,1\rangle$ and $|1,1\rangle$:
\begin{equation}
 S_{new}S_{new}^\dagger =1, \quad S_{new}^\dagger S_{new} = 1-P_{new}, \quad 
P_{new} = |0,0 \rangle \langle 0,0 |+|1,0 \rangle \langle 1,0 |+|0,1 \rangle \langle 0,1 |+|1,1\rangle \langle 1,1 |.
\end{equation}
The explicit form for the new shift operator is as follows
\begin{equation}\left.
\begin{array}{l}
 S_{new} =  1- \displaystyle{\sum_{n'_1=0}^\infty} |n'_1 , 0\rangle \langle n'_1 , 0| \left(1- \frac{1}{\sqrt{M_1+1}}b_1\right)
- \displaystyle{\sum_{n'_1=0}^\infty} |n'_1 , 1\rangle \langle n'_1 , 1| \left(1- \frac{1}{\sqrt{M_1+1}}b_1\right) \\
 S_{new}^\dagger=  1-  \displaystyle{\sum_{n'_1=0}^\infty}\left(1- b_1^\dagger\frac{1}{\sqrt{M_1+1}}\right) |n'_1 , 0\rangle \langle n'_1 , 0|
-  \displaystyle{\sum_{n'_1=0}^\infty}\left(1- b_1^\dagger\frac{1}{\sqrt{M_1+1}}\right) |n'_1 , 1\rangle \langle n'_1 , 1|
\end{array}\right\rbrace
\end{equation}
The operator $b_a$ (corresponding to $a_a$ ) is defined as in (\ref{expression_b}).
We can check that this new solution  satisfies the ASD condition (\ref{asd_r4}) and the equations of motion (\ref{eom_r4}) 
 (for details see Appendices \ref{new_soln_eom} and \ref{new_soln_asd}).
The topological charge of this new solution can be shown to be 4 times the charge of the usual single ASD instanton:  $Q_{new} = -4$
 (see appendix \ref{topo_charge}).

This solution is different from the usual $k=-4$ instanton  for $U(1)$ gauge group despite the topological charge being the same.
We can understand this by observing  that the difference $q$ between the numbers of $a$'s and $a^\dagger$'s  in the two solutions is not same:
 $q=1$  for the usual ADHM solutions irrespective of its charge and the gauge group, whereas
 $q=2$ for the new solution (coming solely because of the operator $b$ in the expression of $\hat{D}_a$ 
given by (\ref{gauge_field_r4_new})).
(The number of $a$'s in $\Psi$ is equal to that of $a^\dagger$'s in $\Psi^\dagger$ and vice-versa.)

The new solution cannot be reduced to the usual instanton by a unitary transformation and hence represents gauge inequivalent configuration.

If we use the operator $b_a(z_{a+},z_{a-})$ defined in (\ref{translation_operator}), we can construct an instanton solution by 
repeating the steps we have outlined above. This instanton also has charge $-4$ as 
the trace of an operator is invariant under unitary transformation. Then the four complex parameters $z_{1+},z_{2+},z_{1-},z_{2-}$ 
can be thought as characterizing the ``locations'' of four instantons with charge $-1$. It is easy to see that in the coincident limit
 $z_{1+}=z_{2+}=z_{1-}=z_{2-}=0$, we recover (\ref{gauge_field_r4_new}).

We can use the above technique to find a new solution in terms of the generalized Bose operator $b_a^{(p_a)}$:

\begin{eqnarray} \label{instanton_solution_b_pa}
 &&\hat{D}_{a}=\frac{1}{\sqrt{\theta}} S_{new}(\hat{\Phi}_{new})^{-\frac{1}{2}} b^{(p_a)}_a(\hat{\Phi}_{new})^
{\frac{1}{2}} S_{new}^{\dagger}, \nonumber \\
&& \hat{\Phi}_{new}=1+\frac{2}{M}, \quad M=M_1^{(p_1)}+M_2^{(p_2)}, \quad M_a^{(p_a)}=b^{(p_a)\dagger}_ab^{(p_a)}_a \nonumber \\
 &&S_{new} =1-\sum_{i=0}^{p_2-1}\sum_{n'_1=0}^{\infty} |n'_1,i\rangle\langle n'_1,i|\left(1-\frac{1}{\sqrt{M^{(p_1)}_1+1}}b^{(p_1)}_1\right),
\end{eqnarray}
with $S_{new}$ satisfying $S_{new}S_{new}^{\dagger}=1$ and $S_{new}^{\dagger}S_{new}=1-\displaystyle\sum_{i=0,j=0}^{i=p_1-1,j=p_2-1}|i,j\rangle \langle i,j|$.
This solution represents an ASD instanton with charge $Q=-p_1p_2$. 
Again, it is gauge inequivalent to the $k=-p_1p_2$  instanton known in the literature.
We could as well have used a different shift operator given by
\begin{equation}\label{shift_operator_2}
S_{new}^\prime =1-\sum_{i=0}^{p_1-1}\sum_{n'_2=0}^{\infty} |i,n'_2\rangle\langle i,n'_2|\left(1-\frac{1}{\sqrt{M^{(p_2)}_2+1}}b^{(p_2)}_2\right).
\end{equation}
Their actions are given by
\begin{equation}
 S_{new}^\dagger |n_1,n_2\rangle = \left\{
\begin{array}{ll}
 |n_1+p_1,n_2\rangle & {\rm if } \,\, 0\leq n_2 \leq p_2-1 \\
|n_1,n_2\rangle & {\rm if } \,\, n_2 \geq p_2,
\end{array} \right.
\end{equation}
\begin{equation}
 S_{new}^{\prime \dagger} |n_1,n_2\rangle = \left\{
\begin{array}{ll}
 |n_1,n_2+p_2\rangle & {\rm if } \,\, 0\leq n_1 \leq p_1-1 \\
|n_1,n_2\rangle & {\rm if } \,\, n_1 \geq p_1.
\end{array} \right.
\end{equation}
Other multi--instanton solutions can be constructed using the reducible representations involving squeezed operators 
(\ref{reducible_squeeze}). 
Our construction of multi--instantonic solutions using reducible representations of the standard harmonic oscillator algebra
 may also be generalized for 4k-dimensional instatons as discussed in \cite{Ivanova:2005fh, Broedel:2007wc}. 
This exercise will be left as a future work\footnote{We thank T. A. Ivanova for pionting out these references to us.}.

\section{Conclusion}\label{conclusion}
We have described static classical solutions of noncommutative gauge theories in various spacetime dimensions 
and showed that the generalized Bose operators are significant in constructing solutions with higher topological numbers.
While the flux tubes and vortices with higher winding numbers correspond to known solutions, the case of 
multi--instantons is different. The multi--instantons with charge $-p_1p_2$ ($p_1,p_2$ non-negative integers) are 
not gauge equivalent to known solutions. Another significant result of this article is an explicit relation between the instanton number and the 
representation theory labels $p_1$ and $p_2$.

Using the ``translated'' $b$ operators (\ref{translation_operator})
 we have constructed multi-instantons that depend explicitly on $p_1p_2$ complex parameters. While the full moduli space of 
noncommutative multi--instantons is still not well understood, we hope that this identification contributes partially to this question.
 
Though we have only considered a few cases, there is actually a large variety of situations in noncommutative 
gauge theories where generalized Bose operators may be used. 
In particular we expect that this procedure may shade new light on merons, monopoles,
dyons, skyrmions etc. We plan to revisit some of these questions in future.

\section*{Acknowledgement} 
SV would like to thank the High Energy Theory Group at McGill University for financial support.

\section*{Appendices}

\appendix
\section{Determination Of $\bar{A}_\infty$ \label{det_gauge_field}}
$\bar{A}_\infty$ can be determined by multiplying (\ref{eom_sublead}) with $\bar{\Phi}_{\infty}$ from right and using (\ref{eom_lead})
\begin{eqnarray}
 \bar{A}_\infty=-i(a- \Phi_{\infty}a\bar{\Phi}_{\infty}).
\label{expression_gauge_field}
\end{eqnarray}
Then (\ref{witten_vortex}) can be used in (\ref{expression_gauge_field}) to determine $\bar{A}_\infty$.  We know that
\begin{eqnarray}
 a^n a^{\dagger n} &=& 
(N+n)(N+n-1).....(N+1).
\end{eqnarray}
Using this, we can show that
\begin{eqnarray}
 \bar{A}_\infty&=&-i\left(a- \Phi_{\infty}a\bar{\Phi}_{\infty}\right)\\
&=& -i\left(a-a\frac{\sqrt{(N+n)(N+n-1).....(N+1)} }{\sqrt{(N+n-1)(N+n-2).....N}}\right)\\
&=& -ia \frac{\sqrt{N}-\sqrt{N+n}}{\sqrt{N}}\\
&=& -i \frac{1}{\sqrt{N+1}} a\left(\sqrt{N}-\sqrt{N+n}\right).
\end{eqnarray}
Similar technique gives us $\bar{A}^{new}_\infty$ as in (\ref{new_gauge_field}).
\section{Large Distance Behavior \label{l_d_b}}
Let $|\omega\rangle $ be the coherent states of the operator $a$, where $\omega$ is a complex number labeling the states. In the large 
$\omega $ limit, the expectation value $\langle\omega|\Phi_{\infty}| \omega 
\rangle$ and $\langle\omega|\bar{A}_\infty| \omega \rangle$ gives the large distance behavior of the solution:
\begin{eqnarray}
 \langle\omega|\Phi_{\infty}| \omega \rangle &=&\langle\omega| \frac{1}{\sqrt{a^n a^{\dagger n}}}a^n| \omega \rangle  \\
&=& \omega^n \langle\omega|  \frac{1}{\sqrt{(N+n)(N+n-1)....(N+1)}}| \omega \rangle  \\
&\approx& \omega^n \langle\omega|  \frac{1}{(\sqrt{N})^n}| \omega \rangle  \\
&\approx& \omega^n (\langle\omega| \frac{1}{N}| \omega \rangle)^{\frac{n}{2}}\\
 &\approx& \omega^n \frac{1}{(\bar{\omega}\omega)^{\frac{n}{2}}} \\
&\approx& e^{i n\varphi}.
\end{eqnarray}
For the gauge field,
\begin{eqnarray}
 \langle\omega|\bar{A}_\infty| \omega \rangle &=& \langle\omega|-i \frac{1}{\sqrt{N+1}} a(\sqrt{N}-\sqrt{N+n})| \omega \rangle \\
&\approx&-i \langle\omega| \frac{N+1}{\sqrt{N+1}}a (1-(1+\frac{n}{2N})| \omega \rangle\\
&\approx& i \frac{n}{2}\langle\omega|a\frac{1}{N}|\omega \rangle\\
&\approx&  i \frac{n}{2}\langle\omega|\frac{1}{N}a|\omega \rangle\\
&\approx&  i\frac{n}{2\bar{\omega}}.
\end{eqnarray}
Similar thing can done for  $\Phi^{new}_{\infty}$ and $\bar{A}^{new}_\infty$ to get (\ref{large_distance2}).

\section{The New Solution Satisfies The Equation Of Motion \label{new_soln_eom}}
The equations of motion are
\begin{equation}
 [\hat{D}_{x^\beta}, [\hat{D}_{x^\beta},\hat{D}_{x^\alpha}]]=0, \quad \alpha,\beta=1,2,3,4.
\end{equation}
There are four equations for different values of $\alpha$. 
We will show only for $\alpha=1$ and other follow similarly.
For $\alpha=1$ the equation becomes
\begin{equation}
 [\hat{D}_{x^2}, [\hat{D}_{x^2},\hat{D}_{x^1}]]+[\hat{D}_{x^3}, [\hat{D}_{x^3},\hat{D}_{x^1}]]+[\hat{D}_{x^4}, [\hat{D}_{x^4},\hat{D}_{x^1}]]=0.
\end{equation}
In terms of the complex coordinates, 
\begin{equation}
 \left([\hat{D}_1,[\hat{D}_{\bar{1}},\hat{D}_1]]+[\hat{D}_2,[\hat{D}_{\bar{2}},\hat{D}_1]]\right)+\left([\hat{D}_{\bar{1}},[\hat{D}_{\bar{1}},\hat{D}_1]]-
[\hat{D}_{\bar{2}},[\hat{D}_{2},\hat{D}_{\bar{1}}]]\right)=0.
\end{equation}
Now using the explicit expression for the solutions (\ref{gauge_field_r4_new}) we get
\begin{eqnarray}
[\hat{D}_{\bar{1}}, [\hat{D}_{\bar{1}},\hat{D}_1]]&=& S_{new} b_1^\dagger \sqrt{\frac{M+3}{M+1}} \sqrt{\frac{M}{M+2}}
\left( \frac{M+2}{M} \frac{M-1}{M+1} M_1 -  
\frac{M}{M+2} \frac{M+3}{M+1}( M_1+1)\right.\nonumber \\  
&& \left. - \frac{M}{M+2} \frac{M+3}{M+1}( M_1+1)+ \frac{M+1}{M+3} \frac{M+4}{M+2}( M_1+2)\right) S_{new}^\dagger,
\end{eqnarray}
\begin{eqnarray}
 [\hat{D}_{\bar{2}}, [\hat{D}_{2},\hat{D}_{\bar{1}}]]&=& -S_{new} b_1^\dagger \sqrt{\frac{M+3}{M+1}} \sqrt{\frac{M}{M+2}}
\left( \frac{M+2}{M} \frac{M-1}{M+1} M_2 -  
\frac{M}{M+2} \frac{M+3}{M+1}M_2\right. \nonumber \\  
&&\left. - \frac{M}{M+2} \frac{M+3}{M+1}( M_2+1)+ \frac{M+1}{M+3} \frac{M+4}{M+2}( M_2+2)\right) S_{new}^\dagger.
\end{eqnarray}
Adding above two we get
\begin{equation}
 [\hat{D}_{\bar{1}}, [\hat{D}_{\bar{1}},\hat{D}_1]]+[\hat{D}_{\bar{2}}, [\hat{D}_{2},\hat{D}_{\bar{1}}]]=0.
\end{equation}
Similarly we can show $[\hat{D}_{\bar{1}},[\hat{D}_{\bar{1}},\hat{D}_1]]-[\hat{D}_{\bar{2}},[\hat{D}_{2},\hat{D}_{\bar{1}}]]=0$
and hence the equation of motion is satisfied.

\section{The New Solution Satisfies The ASD Condition \label{new_soln_asd}}
The commutator between $\hat{D}_a$ and $\hat{D}_{\bar{a}}$ is
\begin{equation}
[\hat{D}_a , \hat{D}_{\bar{a}}] = -\frac{1}{\theta}S_{new}\left(\frac{M}{M+2}\frac{M+3}{M+1} 
(M_{a}+1)-\frac{M+2}{M}\frac{M-1}{M+1}M_{a}\right)S_{new}^\dagger.
\end{equation}
Summing over $a=1,2$ we get
\begin{equation}
 [\hat{D}_1 , \hat{D}_{\bar{1}}]+[\hat{D}_2 , \hat{D}_{\bar{2}}]= -\frac{1}{\theta}S_{new}\left(
M\frac{M+3}{M+1}-(M+2)\frac{M-1}{M+1}\right)S_{new}^\dagger 
=-\frac{2}{\theta}.
\end{equation}
Therefore, $ F_{1\bar{1}}+F_{2\bar{2}} =\frac{2}{\theta}+ [\hat{D}_1 , \hat{D}_{\bar{1}}]+[\hat{D}_2 , \hat{D}_{\bar{2}}]=0$.
Similarly one can show $F_{12}= F_{\bar{1}\bar{2}}=0$.
\section{The Topological Charge Of Instantons \label{topo_charge}}

The topological charge of the instanton is defined as
\begin{equation}
 Q \propto \theta^2 \,\,Tr\left(F_{1\bar{1}}F_{2\bar{2}}-F_{1\bar{2}}F_{2\bar{1}}\right) \implies Q 
= c\theta^2 \displaystyle{\sum_{n_1,n_2=0}^\infty}\langle n_1,n_2|\left(F_{1\bar{1}}F_{2\bar{2}}-F_{1\bar{2}}F_{2\bar{1}}\right)|n_1,n_2\rangle
\end{equation}
where we denote the constant of proportionality by $c$.

Further, the ASD condition gives
\begin{equation}
 F_{1\bar{1}}F_{2\bar{2}} = -F_{1\bar{1}}^2 = -\left( \frac{1}{\theta^2}+\frac{2}{\theta}\left[\hat{D}_1, \hat{D}_{\bar{1}}\right]
+\left[\hat{D}_1, \hat{D}_{\bar{1}}\right]^2\right).
\end{equation}
\subsection{Usual Single ASD solution}
The solutions (\ref{usual_instanton}) can be simplified to
\begin{equation}
 \hat{D}_{a} = \frac{1}{\sqrt{\theta}}S \sqrt{\frac{N(N+3)}{(N+1)(N+2)}}a_a S^{\dagger}, \quad
 \hat{D}_{\bar{a}} = -\frac{1}{\sqrt{\theta}}S a_a^\dagger \sqrt{\frac{N(N+3)}{(N+1)(N+2)}} S^{\dagger}.
\end{equation}
An arbitrary commutator becomes
\begin{equation} \label{d_comm_alpha_beta}
\left[\hat{D}_{a}, \hat{D}_{\bar{b}}\right] = 
-\frac{4}{\theta}S a_b^\dagger a_a\frac{1}{N(N+1)(N+2)} S^\dagger 
-\frac{\delta_{ab}}{\theta}S \frac{N(N+3)}{(N+1)(N+2)} S^\dagger
\end{equation}
and the components of the field strength (and their products) can be calculated to be
\begin{equation}\label{f11_f22}
 F_{1\bar{1}}F_{2\bar{2}} = -\frac{4}{\theta^2}S\frac{(N_2-N_1)^2}{N^2(N+1)^2(N+2)^2}S^\dagger.
\end{equation}
Again using (\ref{d_comm_alpha_beta}) we get
\begin{equation}
 F_{1\bar{2}} =  -\frac{4}{\theta}Sa_2^\dagger a_1\frac{1}{N(N+1)(N+2)}S^\dagger, \quad
 F_{2\bar{1}} =  -\frac{4}{\theta}Sa_1^\dagger a_2\frac{1}{N(N+1)(N+2)}S^\dagger.
\end{equation}
Multiplying the above two and simplifying gives
\begin{equation}
F_{1\bar{2}} F_{2\bar{1}} = \frac{16}{\theta^2}S\frac{(N_1+1)N_2}{N^2(N+1)^2(N+2)^2} S^\dagger \label{f12_f21}.
\end{equation}
Thus we get
\begin{equation}
 Q = -4c \displaystyle{\sum_{n_1=0}^\infty}\displaystyle{\sum_{n_2=1}^\infty}\frac{(n_1+n_2)^2+4n_2}{(n_1+n_2)^2(n_1+n_2+1)^2(n_1+n_2+2)^2}
-4c \displaystyle{\sum_{n_1=0}^\infty}\frac{1}{(n_1+2)^2(n_1+3)^2}
\end{equation}
\subsection{The New Solution}
Again the solutions (\ref{gauge_field_r4_new}) can be written as
\begin{equation}
 \hat{D}_a = \frac{1}{\sqrt{\theta}}S_{new} \sqrt{\frac{M(M+3)}{(M+1)(M+2)}}b_a S_{new}^{\dagger},
\quad
 \hat{D}_{\bar{a}} = -\frac{1}{\sqrt{\theta}}S_{new} b_a^\dagger \sqrt{\frac{M(M+3)}{(M+1)(M+2)}} S_{new}^{\dagger}.
\end{equation}
Then the commutators evaluate to
\begin{equation}
\left[\hat{D}_a, \hat{D}_{\bar{b}}\right] = 
-\frac{4}{\theta}S_{new} b_b^\dagger b_a\frac{1}{M(M+1)(M+2)} S_{new}^\dagger 
-\frac{\delta_{ab}}{\theta}S_{new} \frac{M(M+3)}{(M+1)(M+2)} S_{new}^\dagger.
\end{equation}
The product of the field strength becomes
\begin{equation}\label{f11_f22_n}
 F_{1\bar{1}}F_{2\bar{2}} = -\frac{4}{\theta^2}S_{new}\frac{(M_2-M_1)^2}{M^2(M+1)^2(M+2)^2}S_{new}^\dagger,
\quad
F_{1\bar{2}} F_{2\bar{1}} = \frac{16}{\theta^2}S_{new}\frac{(M_1+1)M_2}{M^2(M+1)^2(M+2)^2} S_{new}^\dagger.
\end{equation}
Hence the charge is
\begin{eqnarray}
 Q_{new}= -4c \displaystyle{\sum_{n_1=0}^\infty}\left(\displaystyle{\sum_{n_2=2}^\infty}
\frac{(m_{n_1}+m_{n_2})^2+4m_{n_2}}{(m_{n_1}+m_{n_2})^2(m_{n_1}+m_{n_2}+1)^2(m_{n_1}+m_{n_2}+2)^2}\right.\nonumber\\ 
 \left. + \frac{2}{(m_{n_1}+2)^2(m_{n_1}+3)^2}\right)
\end{eqnarray}
with
\begin{eqnarray}
 m_n = \frac{1}{2}(n-\lambda^-_n),\quad \lambda^-_n = \left\{\begin{array}{lll} 0 \,\,\,\,\,\,;n={\rm even} \\
                                                     \,1 \,\,\,\,\,\,;n={\rm odd} \end{array}\right.
\end{eqnarray}

Now for all even $n$'s we have $m_n=m_{n+1}$. Hence any absolutely convergent series over 
$n_1$ (or $n_2$) whose terms depend only on $m_{n_1}$ and $m_{n_2}$ can be broken into equal sums to give
\begin{equation}
 \displaystyle{\sum_{n_1=0}^\infty} G\left(m_{n_1},m_{n_2}\right)= 2\displaystyle{\sum_{n_1=0,2,4,...}} G\left(m_{n_1},m_{n_2}\right).
\end{equation}
Hence
\begin{eqnarray}
 \displaystyle{\sum_{n_1=0}^\infty}\displaystyle{\sum_{n_2=2}^\infty} F\left(m_{n_1},m_{n_2}\right)
&=& 2\displaystyle{\sum_{n_1=0,2,4,...}}2\displaystyle{\sum_{n_2=2,4,6,...}} F\left(m_{n_1},m_{n_2}\right) \nonumber \\
&=& 4\displaystyle{\sum_{n_1=0,2,4,...}}\displaystyle{\sum_{n_2=2,4,6,...}}F\left(m_{n_1},m_{n_2}\right).
\end{eqnarray}
Again $m_n=\frac{n}{2}$ for all even $n$'s. Thus the charge becomes
\begin{eqnarray}
 Q_{new} &=& 4\left[ -4c \displaystyle{\sum_{n_1=0,2,4,...}}\displaystyle{\sum_{n_2=2,4,6,...}}\frac{\left(\frac{n_1}{2}+\frac{n_2}{2}\right)^2+4\left(\frac{n_2}{2}\right)}{\left(\frac{n_1}{2}+\frac{n_2}{2}\right)^2\left(\frac{n_1}{2}+\frac{n_2}{2}+1\right)^2\left(\frac{n_1}{2}+\frac{n_2}{2}+2\right)^2} \right. \nonumber\\
&& \left. \,\,\,\,\,\,\,\, -4c \displaystyle{\sum_{n_1=0,2,4,...}}\frac{1}{\left(\frac{n_1}{2}+2\right)^2\left(\frac{n_1}{2}+3\right)^2}\right].
\end{eqnarray}
Redefining $\frac{n_1}{2}\rightarrow n_1$ and $\frac{n_2}{2}\rightarrow n_2$, we get
\begin{equation}
 Q_{new} = 4Q.
\end{equation}

 \bibliographystyle{unsrt}

\end{document}